\pdfoutput=1
\documentclass[runningheads]{llncs}

\usepackage[T1]{fontenc}
\usepackage{newtxtext}
\usepackage[varvw]{newtxmath}
\usepackage{graphicx}
\usepackage{booktabs}
\usepackage{xcolor}
\usepackage{xurl}
\usepackage{microtype}
\usepackage{pgfplots}
\pgfplotsset{compat=1.18}
\usepgfplotslibrary{groupplots}
\usepackage[hidelinks]{hyperref}

\definecolor{keywordblue}{HTML}{0072B2}
\definecolor{typedorange}{HTML}{D55E00}

\newcommand{\passhat}[1]{\textrm{pass\textasciicircum#1}}
\newcommand{\taubench}[0]{$\tau$-bench}
\newcommand{\taurec}[0]{$\tau$-Rec}
\newcommand{\tautwobench}[0]{$\tau^2$-bench}

\providecommand{\taurecresult}[2]{\ifcsname taurec:#1:#2\endcsname\csname taurec:#1:#2\endcsname\else\errmessage{missing tau-Rec result #1 (#2)}\fi}

\providecommand{\pvalr}[1]{\ensuremath{\taurecresult{#1}{pr}}}

\providecommand{\pp}[1]{\taurecresult{#1}{pp}}

\providecommand{\elim}[1]{\taurecresult{elim}{#1}}

\providecommand{\missq}[1]{\taurecresult{missq}{#1}}

\providecommand{\lvl}[1]{\taurecresult{#1}{est}}

\providecommand{\summary}[1]{\taurecresult{summary}{#1}}
\providecommand{\art}[1]{\taurecresult{artifact}{#1}}
\providecommand{\run}[1]{\taurecresult{run}{#1}}
\providecommand{\userstat}[2]{\taurecresult{user/#1}{#2}}

\providecommand{\recog}[1]{\taurecresult{recog}{#1}}
\providecommand{\audit}[1]{\taurecresult{audit}{#1}}
\providecommand{\scripted}[1]{\taurecresult{scripted}{#1}}
\providecommand{\retr}[1]{\taurecresult{retrieval}{#1}}
\providecommand{\taudit}[1]{\taurecresult{taskaudit}{#1}}

\providecommand{\posthoc}[1]{\taurecresult{posthoc}{#1}}
\providecommand{\famstat}[1]{\taurecresult{example}{#1}}

\providecommand{\misread}[1]{\taurecresult{misread}{#1}}

\providecommand{\headroom}[1]{\taurecresult{headroom}{#1}}

\providecommand{\hf}[1]{\taurecresult{hiddenfail}{#1}}
\providecommand{\redial}[1]{\taurecresult{redial}{#1}}
\expandafter\def\csname taurec:exploratory/interface/model_averaged/mixed/pass^1:pp\endcsname{13.2}
\expandafter\def\csname taurec:exploratory/interface/without_qwen38_27b_high/pass^1:pp\endcsname{3.3}
\expandafter\def\csname taurec:primary/interface/model_averaged/pass^1:pp\endcsname{7.1}
\expandafter\def\csname taurec:primary/interface/model_averaged/pass^1:pr\endcsname{=0.0005}
\expandafter\def\csname taurec:secondary/disclosure/hidden_minus_volunteer/satisfiable/pass^1:pp\endcsname{69.3}
\expandafter\def\csname taurec:secondary/disclosure/mixed_minus_volunteer/pass^1:pp\endcsname{23.3}
\expandafter\def\csname taurec:secondary/scale/keyword/5k_minus_20k/pass^1:pp\endcsname{4.2}
\expandafter\def\csname taurec:secondary/scale/structured/5k_minus_20k/pass^1:pp\endcsname{0.8}
\expandafter\def\csname taurec:secondary/simulator/sim_glm53_flash_minus_symbolic/pass^1:pp\endcsname{10.2}
\expandafter\def\csname taurec:secondary/simulator/sim_qwen38_27b_minus_symbolic/pass^1:pp\endcsname{7.4}
\expandafter\def\csname taurec:disclosure/hidden/pass^4:est\endcsname{0.099}
\expandafter\def\csname taurec:disclosure/mixed/pass^4:est\endcsname{0.582}
\expandafter\def\csname taurec:disclosure/volunteer/pass^4:est\endcsname{0.955}
\expandafter\def\csname taurec:user/sim_glm53_flash:hidden_value_stated_tolerant\endcsname{1.4}
\expandafter\def\csname taurec:user/symbolic:mixed_all_askable_stated_tolerant\endcsname{84.7}
\expandafter\def\csname taurec:addendum/gpt_users/sim_glm53_flash_minus_symbolic/pass^1:pp\endcsname{1.2}
\expandafter\def\csname taurec:addendum/prompt_variant/gpt61_sol_high/mixed/pass^1:pp\endcsname{59.0}
\expandafter\def\csname taurec:ablation/deepseek_v41_flash_high/field_list/all/pass^1:pp\endcsname{4.6}
\expandafter\def\csname taurec:ablation/deepseek_v41_flash_high/retrieval_with_field_list/all/pass^1:pp\endcsname{1.4}
\expandafter\def\csname taurec:ablation/qwen38_27b_high/field_list/all/pass^1:pp\endcsname{0.2}
\expandafter\def\csname taurec:ablation/qwen38_27b_high/retrieval_with_field_list/all/pass^1:pp\endcsname{19.4}
\expandafter\def\csname taurec:summary:final_recommendations\endcsname{24,644}
\expandafter\def\csname taurec:summary:gpt_complete_n\endcsname{101}
\expandafter\def\csname taurec:summary:gpt_no_elicitation_hi\endcsname{49}
\expandafter\def\csname taurec:summary:gpt_no_elicitation_lo\endcsname{40}
\expandafter\def\csname taurec:summary:hidden_nvr\endcsname{0.047}
\expandafter\def\csname taurec:summary:open_no_elicitation_hi\endcsname{25}
\expandafter\def\csname taurec:summary:unseen_recommendations\endcsname{3}
\expandafter\def\csname taurec:run:trials_all\endcsname{32,976}
\expandafter\def\csname taurec:artifact:constraints_complex\endcsname{5}
\expandafter\def\csname taurec:artifact:constraints_medium\endcsname{3--4}
\expandafter\def\csname taurec:artifact:constraints_simple\endcsname{2}
\expandafter\def\csname taurec:artifact:countries\endcsname{166}
\expandafter\def\csname taurec:artifact:country_predicates\endcsname{68}
\expandafter\def\csname taurec:artifact:country_predicates_us_uk\endcsname{51}
\expandafter\def\csname taurec:artifact:eligible_pool\endcsname{26.1}
\expandafter\def\csname taurec:artifact:english_pct\endcsname{83.2}
\expandafter\def\csname taurec:artifact:field_sets\endcsname{13}
\expandafter\def\csname taurec:artifact:format_families\endcsname{36}
\expandafter\def\csname taurec:artifact:formats\endcsname{audiobook, ebook, hardcover, paperback}
\expandafter\def\csname taurec:artifact:genres\endcsname{19}
\expandafter\def\csname taurec:artifact:hidden_selectivity_pct\endcsname{50}
\expandafter\def\csname taurec:artifact:history_families\endcsname{48}
\expandafter\def\csname taurec:artifact:history_two_hidden\endcsname{17}
\expandafter\def\csname taurec:artifact:languages\endcsname{151}
\expandafter\def\csname taurec:artifact:median_answers_20000\endcsname{40}
\expandafter\def\csname taurec:artifact:median_answers_5000\endcsname{10}
\expandafter\def\csname taurec:artifact:nvr_witness_min\endcsname{3}
\expandafter\def\csname taurec:artifact:overrides\endcsname{25}
\expandafter\def\csname taurec:artifact:source_responses\endcsname{21}
\expandafter\def\csname taurec:artifact:track_id\endcsname{books-\allowbreak{}wikidata-\allowbreak{}v2}
\expandafter\def\csname taurec:artifact:uk_pct\endcsname{13.4}
\expandafter\def\csname taurec:artifact:us_pct\endcsname{41.5}
\expandafter\def\csname taurec:artifact:year_max\endcsname{2026}
\expandafter\def\csname taurec:artifact:year_min\endcsname{1460}
\expandafter\def\csname taurec:recog:field_p\endcsname{0.898}
\expandafter\def\csname taurec:recog:field_r\endcsname{0.944}
\expandafter\def\csname taurec:recog:generic_r\endcsname{0.77}
\expandafter\def\csname taurec:recog:turns\endcsname{200}
\expandafter\def\csname taurec:audit:agreed\endcsname{297}
\expandafter\def\csname taurec:audit:judgments\endcsname{300}
\expandafter\def\csname taurec:audit:overrides_sampled\endcsname{12}
\expandafter\def\csname taurec:audit:records\endcsname{60}
\expandafter\def\csname taurec:taskaudit:families\endcsname{30}
\expandafter\def\csname taurec:taskaudit:judgments\endcsname{150}
\expandafter\def\csname taurec:taskaudit:nvr\endcsname{15}
\expandafter\def\csname taurec:scripted:hidden_nvr_solved\endcsname{2}
\expandafter\def\csname taurec:scripted:hidden_nvr_total\endcsname{15}
\expandafter\def\csname taurec:scripted:hidden_sat_panel\endcsname{45}
\expandafter\def\csname taurec:scripted:hidden_sat_panel_solved\endcsname{28}
\expandafter\def\csname taurec:scripted:hidden_sat_solved\endcsname{84}
\expandafter\def\csname taurec:scripted:hidden_sat_total\endcsname{135}
\expandafter\def\csname taurec:scripted:hidden_satisfiable_rel\endcsname{0.600}
\expandafter\def\csname taurec:scripted:open_total\endcsname{300}
\expandafter\def\csname taurec:elim:hidden_nvr_panel\endcsname{6}
\expandafter\def\csname taurec:elim:hidden_nvr_panel_solved\endcsname{5}
\expandafter\def\csname taurec:elim:hidden_nvr_solved\endcsname{13}
\expandafter\def\csname taurec:elim:hidden_sat_median_probes\endcsname{3}
\expandafter\def\csname taurec:elim:hidden_sat_panel_solved\endcsname{43}
\expandafter\def\csname taurec:elim:hidden_sat_solved\endcsname{129}
\expandafter\def\csname taurec:elim:hidden_satisfiable_rel\endcsname{1.000}
\expandafter\def\csname taurec:retrieval:keyword_20000\endcsname{73}
\expandafter\def\csname taurec:retrieval:keyword_5000\endcsname{106}
\expandafter\def\csname taurec:posthoc:capped3_all\endcsname{0.208}
\expandafter\def\csname taurec:posthoc:capped3_eliminating\endcsname{0.622}
\expandafter\def\csname taurec:posthoc:capped3_eliminating_rel\endcsname{0.733}
\expandafter\def\csname taurec:posthoc:capped3_scripted\endcsname{0.311}
\expandafter\def\csname taurec:posthoc:hidden_complex\endcsname{0.163}
\expandafter\def\csname taurec:posthoc:hidden_satisfiable\endcsname{0.283}
\expandafter\def\csname taurec:posthoc:hidden_simple\endcsname{0.409}
\expandafter\def\csname taurec:posthoc:leak_tasks\endcsname{5}
\expandafter\def\csname taurec:posthoc:mde_pp\endcsname{6}
\expandafter\def\csname taurec:posthoc:mixed_complete_success_pct\endcsname{98}
\expandafter\def\csname taurec:posthoc:mixed_incomplete_success_pct\endcsname{52}
\expandafter\def\csname taurec:posthoc:mixed_satisfiable\endcsname{0.756}
\expandafter\def\csname taurec:posthoc:pairs_hidden_mde_pp\endcsname{14}
\expandafter\def\csname taurec:posthoc:pairs_mde_pp\endcsname{8}
\expandafter\def\csname taurec:posthoc:pairs_separated\endcsname{19}
\expandafter\def\csname taurec:posthoc:pairs_total\endcsname{28}
\expandafter\def\csname taurec:posthoc:run_usd_hi\endcsname{4.4}
\expandafter\def\csname taurec:posthoc:run_usd_lo\endcsname{1.0}
\expandafter\def\csname taurec:posthoc:signflip_template_primary\endcsname{0.15}
\expandafter\def\csname taurec:posthoc:volunteer_satisfiable\endcsname{0.975}
\expandafter\def\csname taurec:misread:drop_all_pp\endcsname{3.9}
\expandafter\def\csname taurec:missq:labelled_pp\endcsname{2.0}
\expandafter\def\csname taurec:redial:after_suggestion_pct\endcsname{12}
\expandafter\def\csname taurec:redial:answered_question_pct\endcsname{52}
\expandafter\def\csname taurec:redial:dialogues\endcsname{11,348}
\expandafter\def\csname taurec:redial:unprompted_pct\endcsname{36}
\expandafter\def\csname taurec:hiddenfail:gpt_failures\endcsname{416}
\expandafter\def\csname taurec:hiddenfail:gpt_untested\endcsname{415}
\expandafter\def\csname taurec:hiddenfail:refused_untested_pct\endcsname{73}
\expandafter\def\csname taurec:hiddenfail:untagged_acceptable_pct\endcsname{8.5}
\expandafter\def\csname taurec:hiddenfail:untested_pct\endcsname{78}
\expandafter\def\csname taurec:headroom:ask_mixed_lo\endcsname{0.931}
\expandafter\def\csname taurec:headroom:protocol_capped_hi\endcsname{0.600}
\expandafter\def\csname taurec:headroom:protocol_capped_lo\endcsname{0.383}
\expandafter\def\csname taurec:headroom:protocol_gain_pp\endcsname{22.5}
\expandafter\def\csname taurec:headroom:protocol_nvr_hi_pct\endcsname{62}
\expandafter\def\csname taurec:headroom:protocol_nvr_lo_pct\endcsname{46}
\expandafter\def\csname taurec:headroom:protocol_sat_hi\endcsname{0.825}
\expandafter\def\csname taurec:headroom:protocol_sat_lo\endcsname{0.708}
\expandafter\def\csname taurec:headroom:protocol_vs_popularity_above\endcsname{3}
\expandafter\def\csname taurec:headroom:verify_nvr_hi_pct\endcsname{8}
\expandafter\def\csname taurec:headroom:verify_sat_hi\endcsname{0.617}
\expandafter\def\csname taurec:headroom:verify_sat_lo\endcsname{0.483}
\expandafter\def\csname taurec:headroom:verify_untested_hi_pct\endcsname{30}
\expandafter\def\csname taurec:example:answers\endcsname{27}

\newcommand{\primary}{primary/interface/model_averaged/pass^1}
\newcommand{\gptvariant}{addendum/prompt_variant/gpt61_sol_high}

\newcommand{\dsabl}{ablation/deepseek_v41_flash_high}
\newcommand{\qwabl}{ablation/qwen38_27b_high}
\hypersetup{
  pdftitle={τ-Rec-Books: A Disclosure-Controlled Benchmark for Requirement Elicitation in Agentic Recommender Systems},
  pdfauthor={Bharath Sivaram Narasimhan; Karthik R. Narasimhan},
  pdfsubject={A disclosure-controlled books benchmark for requirement elicitation in agentic recommender systems},
  pdfkeywords={Conversational recommendation, Verifiable evaluation, Preference elicitation, LLM agents, User simulation}
}

\begin{document}
\raggedbottom
\setlength{\emergencystretch}{1.5em}

\title{\taurec{}-Books: A Disclosure-Controlled Benchmark for Requirement
Elicitation in Agentic Recommender Systems}
\titlerunning{\taurec{}-Books: A Disclosure-Controlled Benchmark}

\author{Bharath Sivaram Narasimhan\inst{1}
\and Karthik R. Narasimhan\inst{2}}
\authorrunning{B. S. Narasimhan and K. R. Narasimhan}

\institute{Independent Researcher, Mountain View, CA, USA\\
\email{nbharaths@gmail.com}
\and
Princeton University, Princeton, NJ, USA\\
\email{karthikn@princeton.edu}}

\maketitle

\begin{abstract}
While conversational recommendation systems provide a more natural user interface, they face the challenge of progressively discovering a user's needs before they can recommend an item that satisfies them.  To evaluate the ability of AI agents at this elicitation process, we introduce \taurec{}-Books, a benchmark that holds each user request fixed while varying only \textit{when} the user reveals their requirements: all at the start (\textit{volunteer}), one at the start and the rest when asked (\textit{mixed}), or some never (\textit{hidden}).  A hidden requirement can be found only by proposing books and inferring it from the user's accept or reject verdicts.  We generate 450 tasks over a catalog of 20,000 books from Wikidata, develop a rule-based simulated user to answer the agent, and score recommendations against catalog metadata without an LLM judge.  Across four agents and two search interfaces, success rates fall for every agent and interface, on
average from \posthoc{volunteer_satisfiable} (volunteer) to
\posthoc{mixed_satisfiable} (mixed) and \posthoc{hidden_satisfiable} (hidden).
Mixed tasks fail mainly because agents do not ask, and when told to ask, every agent reaches at least \headroom{ask_mixed_lo}.  On hidden tasks, most failures come from agents recommending a book they never proposed to the user.  When told to propose first, agents only match a simple baseline that proposes popular books, and stay well below one that uses rejection to rule out possible requirements. We release the catalog pipeline, task generator, simulated user, validators,
reference agents, and all agent traces at
{\def\UrlBreaks{\do\/}\url{https://github.com/nbharaths/tau-rec/tree/main/data/tracks/books-wikidata-v2}}.

\end{abstract}

\section{Introduction}

Agentic recommenders combine dialogue with catalog search and metadata tools,
so their evaluation must test more than ranking quality.
\taurec{} introduced verifiable rewards for this setting, scoring recommendations
against catalog predicates without an LLM judge \cite{narasimhan2026taurec}.  On a small movie catalog, success fell steeply as requirements moved from stated to hidden.

However, \taurec{} leaves open a few questions around different conditions that may impact agent success: (1) the timing of information disclosure given a fixed request, (2) the structure of the search interface, and (3) the overall size of the item catalog.
The first question matters greatly because real users rarely state every requirement up front.  For instance, in ReDial's \redial{dialogues} human
recommendation dialogues \cite{li2018redial}, users state only
\redial{unprompted_pct}\% of their first genre, era, or origin requirements
unprompted, \redial{answered_question_pct}\% in answer to a question, and
\redial{after_suggestion_pct}\% only after a suggestion.
Users of a natural-language recommender also often
add criteria after seeing results \cite{kang2017understanding}.

We introduce \taurec{}-Books, a benchmark that makes disclosure
a controlled variable.  Each task family fixes a request (its predicates,
answer set, persona, and reading history) and comes in a volunteer, a mixed,
and a hidden variant that differ only in \textit{when} each requirement is revealed.
A rule-based simulated user enforces the disclosure
rules, and the scoring does not make use of LLMs.  Agents can find books through one of two search interfaces: keyword search, or structured search, which filters on catalog fields.

In our experiments, we find that success
rate on satisfiable tasks falls for every model and interface (across 4 models, 2 interfaces) from
\posthoc{volunteer_satisfiable} on volunteer to \posthoc{mixed_satisfiable} on
mixed and \posthoc{hidden_satisfiable} on hidden tasks.  Moreover, structured search beats keyword search by \pp{\primary} points on average. In our error analysis, we find that mixed failures come from not asking and hidden failures from recommending a book without first proposing it to the user.  We also show that when agents are told to propose first, they match a scripted agent that proposes popular books, but not one that also reads declines and rejections.  Overall, we
contribute (i) 450 tasks in matched disclosure variants, generated from catalog predicates without a retriever; (ii) a frozen, audited 20,000-work public-domain (CC0) catalog with nested 5k and 10k
subsets; and (iii) validators, LLM-free reference agents, and the traces of all \run{trials_all} trials, which
regenerate every result.

\section{Related Work}

\textbf{Corpora, test collections, and toolkits.}  Conversational recommender
system (CRS) corpora range from human--human movie conversations
\cite{li2018redial,radlinski2019ccpe} to knowledge-grounded
\cite{moon2019opendialkg} and synthesized ones \cite{kim2024pearl}.  They support
training and replay but not counterfactual conditions \cite{jannach2023evaluating},
and CRS toolkits prescribe no executable task state \cite{zhou2021crslab}.  For
books, Goodreads supplies ratings and reviews and Book-Crossing ratings
\cite{wan2018goodreads,ziegler2005bookcrossing}.  Social Book Search collected
complex reader requests \cite{koolen2015looking}, later formalized as
narrative-driven recommendation \cite{bogers2017narrative}.  KITAB and QUEST test
constraint satisfaction and set queries over entities that include books
\cite{abdin2024kitab,malaviya2023quest}.

\textbf{Eliciting requirements.}  Multi-round CRSs ask about items
\cite{christakopoulou2016towards} or attributes.  Attribute-based CRSs learn which
attribute to ask about and when to recommend \cite{sun2018conversational,lei2020ear}.
Their simulated users hold a target item, answer every attribute question, and
reject wrong items without reasons.  Our mixed and hidden conditions pose the same
decision with free questions and verifiable outcomes, and our hidden users also
decline some questions.  Searching by verdicts resembles twenty-questions games for
LLMs \cite{hu2024uot} and LLM preference elicitation \cite{austin2024pebol}.
Constraint-based recommenders relax queries that return nothing
\cite{mcsherry2004incremental}, while we score abstention.

\textbf{Agentic recommendation.}  LLMs act as conversational recommenders
\cite{he2023zeroshot} and increasingly use tools \cite{wang2024recmind}.
InteRecAgent's SQL-style and similarity retrieval \cite{huang2025interecagent} is the
closest precedent for our interface contrast.  AgentRecBench scores tool-using
recommender agents by hit rate, with no user in the loop
\cite{shang2025agentrecbench}.  WebShop scores attribute-matched product choices
\cite{yao2022webshop}.  In EComAgentBench, requirements held in a user profile or revealed only through
clarification are met less often than those stated in the query
\cite{du2026ecomagentbench}.  \taubench{}, ToolSandbox, and \tautwobench{} established
stateful tool evaluation \cite{yao2025taubench,lu2025toolsandbox,barres2025tautwo},
which \taurec{} brought to recommendation \cite{narasimhan2026taurec}.

\textbf{Clarifying questions.}  Clarifying questions are well studied
in information retrieval \cite{aliannejadi2019qulac}, as are simulated users that
answer them \cite{sekulic2022usi}.  LLMs degrade when a request is spread over turns,
and they rarely ask for what is missing \cite{laban2026lost,li2024mediq}.  UserBench
users reveal preferences gradually \cite{qian2025userbench}.  Other benchmarks
test asking about implicit intentions \cite{qian2024tellmemore}, choosing the one
question that recovers a missing variable \cite{li2025questbench}, and asking about
hard constraints before planning \cite{zhang2024askbeforeplan}.  LHAW removes details
from long-horizon tasks and lets agents ask a simulated user \cite{pu2026lhaw}.
COMPASS varies how fast a simulated traveller reveals hard constraints
\cite{qin2025compass}.  RegretBench scores clarification policies by regret,
including in product recommendation \cite{ta2026regretbench}.  Ambig-SWE gives coding
agents full, underspecified, or askable versions of the same issue
\cite{vijayvargiya2026ambigswe}, the closest precedent for matched disclosure
variants.  We add a condition where the user declines to state some requirements but
judges proposals, and we vary catalog size and interface on the same tasks.

\textbf{User simulation.}  Agenda-based simulators drive task-oriented
dialogue evaluation \cite{schatzmann2007agenda}, including database-checked success
on MultiWOZ \cite{budzianowski2018multiwoz}, and CRS evaluation
\cite{zhang2020evaluating}.  UserSimCRS v2 packages agenda-based and LLM simulators
\cite{bernard2026usersimcrs}.  LLM users can evaluate CRSs \cite{wang2023ievalm}, but
they reveal what they should withhold and express preferences unlike people
\cite{yoon2024simulators,zhu2024reliable}.  Modular, human-validated, and
crowdsourced studies address these gaps
\cite{chen2025recusersim,meshi2026convapparel,bernard2025crsarena}, and Sim4IA-Bench
scores simulated queries and utterances against real search sessions
\cite{kruff2026sim4ia}.  Petruzzelli et al.\ vary user stereotypes
\cite{petruzzelli2026stereotypes}; we fix the facts and vary when they are disclosed.

\section{The Books Resource}
\label{sec:conditions}

\textbf{Catalog.}  We source works for the catalog from Wikidata \cite{vrandecic2014wikidata}.  We keep works that Wikidata lists as literary or written works with an author,
publication date, language, English title, and mapped genre, and filter out
editions, series, articles, chapters, plays, poems, and short stories.  We deduplicate records by normalized title and authors, shuffle them in a fixed random order, and take the first 20,000 of the \art{eligible_pool}k eligible works to form the catalog; the 5k and 10k catalogs are the first 5,000 and 10,000 of these, so each smaller catalog lies inside the larger ones.  We correct
\art{overrides} Wikidata labels and drop two kinds of faulty records
(Sect.~\ref{sec:audit}).  The
catalog covers \art{genres} genres, \art{languages} language and
\art{countries} country labels, and years \art{year_min}--\art{year_max}.

\begin{figure}[tbp]
\centering
\fbox{\begin{minipage}{0.96\linewidth}\scriptsize\raggedright
\setlength{\tabcolsep}{3pt}%
\def\turn#1{\par\hangindent=3em\hangafter=1\noindent
  \makebox[3em][l]{\textit{#1}}\ignorespaces}%
\def\cond#1{\makebox[4em]{\textit{#1}}}%
\def\stated{\textcolor{black!60}{stated}}%
\begin{minipage}[t]{0.38\linewidth}\raggedright
\textbf{Family 060} (medium, \famstat{answers} answers at 20k).\\
The mixed and hidden users open with ``I'm looking for one book.  It should
include Science Fiction.''
\end{minipage}\hfill
\begin{tabular}[t]{@{}lccc@{}}
\textit{Predicate} & \cond{Volunteer} & \cond{Mixed} & \cond{Hidden}\\
genre includes Science Fiction & \stated & \stated & \stated\\
genre includes Young Adult & \stated & on ask & hidden\\
published no later than 2001 & \stated & on ask & on ask
\end{tabular}\par\vspace{1.4pt}
\textbf{Mixed} (DeepSeek, structured; success).
\turn{Agent}``\ldots{} Any particular language \ldots?  \ldots{} A publication
year range \ldots?  \ldots{} Any additional genres \ldots?''
\turn{User}``\ldots{} it should include Young Adult; and it should have been
published no later than 2001.  \ldots''

\turn{Agent}``\ldots{} Have Space Suit---Will Travel by Robert A.\ Heinlein
(1958) \texttt{[[candidate\_id:wd\_Q1213656]]} \ldots''
\turn{User}``\#\#\#ACCEPTED\#\#\# \ldots''\quad\textit{Tool}\enspace
\texttt{recommend(item\_id=wd\_Q1213656)} succeeds.\par\vspace{1.6pt}
\textbf{Hidden} (DeepSeek, structured; failure), after a similar question.
\turn{User}``\ldots{} it should have been published no later than 2001.  I'd
rather not say more about the genre.  \ldots''
\turn{Agent}after 93 searches proposes \emph{I, Robot} (rejected), then recommends
\emph{Fahrenheit 451} without proposing it: science fiction, 1953, not Young Adult.
\end{minipage}}
\caption{Family 060 in its three disclosure conditions, with two trial
excerpts (\ldots{} omits text).}
\label{fig:example}
\end{figure}

\textbf{Task families.}  The 450 tasks are divided into 150 task families, each of which fixes its predicates, answer set,
persona, reading history and accepted formats, and varies only how its
requirements are disclosed (Fig.~\ref{fig:example}).  The predicates use four
fields: genre, language, country of origin, and publication year.  We split the task families into three tiers -- simple, medium, and complex -- with
\art{constraints_simple}, \art{constraints_medium}, and
\art{constraints_complex} predicates, respectively (a year range counts as two), and together follow
\art{field_sets} templates, each a fixed combination of fields and year bounds.  In the fifteen families with no valid recommendation (NVR), dropping any one
predicate leaves at least \art{nvr_witness_min} matches.  Every predicate is
material (i.e., dropping it adds at least one item, and at least 5\%, to the answer set), and
every satisfiable family keeps an acceptable answer (unread, in an accepted
format) in the 5k catalog.

\textbf{Reveal tags and conditions.}  Each predicate carries a reveal tag
(stated, on ask, or hidden), and the three conditions differ
only in these tags.  A \emph{volunteer} user opens with a fully formulated
request that states every predicate and its accepted formats.  A \emph{mixed}
user states one predicate and reveals an on-ask predicate, or its accepted
formats, when asked about the field, so one question naming all fields discloses
every askable fact.  A
\emph{hidden} user also withholds at least one predicate for the whole dialogue;
asked about its field, it declines, which reveals that a requirement exists but
not its value (Table~\ref{tab:usercard} lists every reply).  Together, the hidden predicates rule out at least \art{hidden_selectivity_pct}\% of the books that
match the revealed ones, and no revealed fact gives a hidden one away: a task never reveals
British English, say, while hiding the United Kingdom as the country.

\begin{table}[tbp]
\caption{The rule-based user's replies and the agents' system prompt (verbatim
excerpts).}
\label{tab:usercard}
\centering
\scriptsize
\setlength{\tabcolsep}{4pt}
\begin{tabular}{@{}p{0.17\linewidth}p{0.79\linewidth}@{}}
\toprule
Agent turn & Rule-based user's reply \\
\midrule
Field question & Its on-ask facts; ``I'd rather not say (more) about'' a field
holding a hidden fact; ``No preference on'' any other field \\
``Anything else?'' & One undisclosed on-ask fact; once none remains, ``That's everything I can share'' \\
One tagged book & \texttt{\#\#\#ACCEPTED\#\#\#} if it meets every predicate, is
unread, and exists in any format the user named; else \texttt{\#\#\#REJECTED\#\#\#},
never with a reason \\
Several tags, or none & No verdict; ``Please choose one catalog title \ldots'', or,
if it recognizes no question, ``Please ask about anything you need \ldots'' \\
\midrule
Prompt & Agent instruction \\
\midrule
Elicitation & ``Engage the user in natural conversation to understand their
preferences \ldots'' \\
Proposals & ``\ldots{} append its exact catalog ID as [[candidate\_id:ID]] \ldots'' \\
Verdicts & ``\ldots{} Use rejections as feedback to pick a better candidate, then
call recommend() with your updated pick.'' \\
Ask & adds ``The user may have requirements they have not stated yet; ask about
them before you recommend a book.'' \\
Ask-and-verify & also adds ``Before you call recommend, propose the book you intend
to recommend, with its candidate marker, and wait for the user's verdict.'' \\
Protocol & also adds ``If the user would rather not say (more) about a field, they
have a requirement on that field that they will not state.  A rejected book fails at
least one of the user's requirements.'' \\
\bottomrule
\end{tabular}
\end{table}

\textbf{A trial.}  The agent asks questions, may propose books for verdicts,
and ends with the terminal \texttt{recommend} tool, called with one book or, to abstain, with none.
A trial succeeds when its ending is correct (a book meeting every predicate on
a satisfiable task, an abstention on an NVR task) and the protocol checks hold.

\textbf{What each condition measures.}  Volunteer and mixed tasks separate
satisfying stated requirements from deciding to elicit them (one broad question
suffices), following needs revealed over a conversation
\cite{radlinski2017theoretical}.  Hidden tasks are a stress test.
Real users sometimes state a requirement only after seeing a suggestion; that is
critiquing \cite{chen2012critiquing}, which we leave to later work.  The hidden user
never states it, so the agent must propose books before committing and search using
verdicts that give no values or reasons.

\textbf{Formats, histories, and checks.}  Wikidata does not record which formats a
book comes in, so each book gets a simulated but fixed set (\art{formats}), drawn from a hash
of its Wikidata ID.  In \art{format_families} families the user accepts only one format, and
at least one otherwise valid book in the 5k catalog lacks it.  In \art{history_families} families the user has a reading history that holds one valid answer and, for each hidden
predicate, a book that fails only that predicate, so no single book in it gives a hidden value
away.  In the \art{history_two_hidden} of these with two hidden predicates, the history as a
whole hints at them, but agents do no better on the \posthoc{leak_tasks} such tasks in the
\emph{panel}, the random 51 of the 150 families that every agent runs
(Table~\ref{tab:design}).  Personas are synthetic, and only LLM users voice them.  Besides
meeting the predicates, a trial must pass five protocol checks: the agent recommends at most
one book, never one from the reading history, checks its formats once the user names one,
abstains on NVR tasks, and calls \texttt{recommend} on the others.

\section{Controlled Evaluation Design}
\label{sec:design}

We ask four research questions:  \textbf{RQ1 (disclosure):} how much success
is lost as requirements move from volunteered to askable or hidden, with the
request held fixed?  \textbf{RQ2 (interface):} how does a schema-aligned
search interface change end-to-end success over top-20 keyword search?  \textbf{RQ3 (scale):} how does success change with catalog size when every
5k answer is retained?
\textbf{RQ4 (reliability):} which effects persist across models, trials, and users?

\subsection{Experimental Setup}

\textbf{Agents.}  The agents run on three open-weight models -- DeepSeek V4.1 Flash,
GLM 5.3 Flash, and Qwen 3.8 27B, at temperature 0
(outputs still vary) -- and the frontier model GPT-6.1 Sol at
its required default temperature, all with high reasoning effort.  Every agent gets the same system prompt
(Table~\ref{tab:usercard}), which says that recommending a book that misses any
requirement fails, and so does abstaining when the catalog has an answer. 

\textbf{Interfaces.}  Both interfaces offer metadata, format-availability, and
reading-history tools and the terminal \texttt{recommend(item\_id)}, which scores
the identifier passed, never a title parsed from text.  The \emph{keyword} interface sums per-field Okapi BM25
scores \cite{robertson2009probabilistic} and returns the top 20 without paging;
the year field is not indexed, so year ranges are inexpressible.  The \emph{structured} interface offers
case-insensitive conjunctive filters on genre, language, country (a parent
label covers its children), year range, and acceptable formats, returning pages
of 20 by Wikidata sitelink count (a popularity proxy) or recency; an unknown value
returns the valid vocabulary.  The structured tool's prompt line lists these fields and the
keyword line does not, so the contrast compares two bundles of tool and prompt
over the same metadata, not retrievers alone.

\textbf{Users.}  By default the user is rule-based: it holds the
task facts and answers by fixed rules (Table~\ref{tab:usercard}), using keyword
matching to tell which fields each agent turn asks about.  We checked the matching
against an LLM's labels on \recog{turns} held-out agent turns:
precision \recog{field_p}, recall \recog{field_r}, and recall \recog{generic_r} on
generic requests such as ``Anything else?''.  In the
LLM-user block, GLM 5.3 Flash and Qwen 3.8 27B play the user instead, at
temperature~1: they get
the same rules, facts, and reveal tags but decide what to disclose and word their
own replies, while verdicts still come from the identifier check.  The base prompt
tells agents to use rejections to pick a better candidate, but not that a
rejection means a requirement failed or that a declined question hides one.  A
trial allows 20 agent turns, including turns that draw no verdict, and 600\,s of
agent time, and also ends at a context overflow or after ten calls past a budget
of 150 tool calls in one turn; agents are not told these caps.

\textbf{Trials.}  Each panel task runs four times, and the 18 families of its
\emph{reliability block} run eight times (Table~\ref{tab:design}).

\begin{table}[tbp]
\caption{Experimental blocks; the panel is the reliability block plus \emph{Panel, other}.  Blocks use the rule-based user and the 20k catalog unless a row says otherwise; 32,976 trials in all.}
\label{tab:design}
\centering
\scriptsize
\setlength{\tabcolsep}{2.5pt}
\begin{tabular}{@{}lllllr@{}}
\toprule
Block & Agents & Interfaces & Conditions & Families, trials/task & RQ \\
\midrule
Panel, other & all four & both & all three & 33, 4 & 1, 2, 4 \\
Reliability block & all four & both & all three & 18, 8 & 1, 2, 4 \\
Scale study (5k, 10k, 20k) & open-weight & both & not hidden & 135, 4 & 3, 4 \\
LLM users & DeepSeek & structured & all three & 18, 4 $\times$ 2 users & 4 \\
Ask prompt & all four & structured & all three & 18, 8 & 1 \\
LLM users & GPT-6.1 Sol & structured & all three & 18, 4 $\times$ 2 users & 4 \\
Field list & DeepSeek, Qwen & keyword & all three & 18, 8 & 2 \\
Verify prompt & all four & structured & all three & 18, 8 & 1 \\
Protocol prompt & all four & structured & all three & 18, 8 & 1 \\
\bottomrule
\end{tabular}
\end{table}

\subsection{Outcome Measures and Statistical Analysis}

\textbf{Outcome and uncertainty.}  With \(c_t\) successes among
\(n_t\) trials of task \(t\) in the evaluated set \(T\),
\(\passhat{k}=\frac{1}{|T|}\sum_{t\in T}\binom{c_t}{k}/\binom{n_t}{k}\).
Intervals and $p$ values come from a paired bootstrap that resamples task
families \cite{sakai2006bootstrap,smucker2007comparison};
differences are in points (hundredths of $\passhat{1}$).

\textbf{Tests.}  We run twelve tests;
all other analyses are exploratory.  The \emph{interface test}, structured minus
keyword search at 20k averaged over agents and conditions, is tested on its
own, without correction, at \(p<0.05\).  The other eleven use Holm's correction for multiple comparisons
\cite{holm1979simple}: the interface test per agent, the two disclosure gaps (mixed
and hidden against volunteer), the two scale effects (5k against 20k), their
interaction with the interface, and the two LLM users against the rule-based one.
We also run each interface test without trials that end in a context overflow or at
the tool budget and, as a robustness check, test each contrast across templates with an
exact sign-flip test \cite{smucker2007comparison}.

\section{Resource Validation}
\label{sec:census}\label{sec:audit}

\textbf{Automated checks.}  A validator recomputes every answer set and
acceptable answer, rechecks the NVR families and the nested catalogs, and runs the
structured tool on every satisfiable family and catalog size without error.

\textbf{Reference agents.}  The two scripted agents
run through the same user, tools, scoring, and caps as the LLM agents, and
propose only unread books.  The popularity agent asks one question naming every field, filters on what the user
states, and proposes the most popular match until one is accepted, abstaining when
none is left; it uses only what the prompt tells agents.  The eliminating agent adds the protocol's meaning: a
declined field holds a requirement, and a rejected book fails at least one.  It
lists the possible hidden requirements (one or two values of each declined
field), drops every one a rejected book satisfies, proposes the book that fits the most of
those left, and abstains when no candidate fits any.  Both solve all
\scripted{open_total} volunteer and mixed tasks.  Of the
\scripted{hidden_sat_total} satisfiable and \scripted{hidden_nvr_total} NVR hidden
tasks, the popularity agent solves \scripted{hidden_sat_solved} and
\scripted{hidden_nvr_solved}, the eliminating agent \elim{hidden_sat_solved} and
\elim{hidden_nvr_solved} (median \elim{hidden_sat_median_probes} proposals per
success).  Both parse the rule-based user's reply templates, so
they test solvability, not language handling.  A no-ask floor never asks and recommends
the most popular unread book that matches the opening request.

\textbf{Catalog audit.}  Two annotators separately reviewed
\audit{records} records (30 across genres, 30 among the 500
most popular) under written rules and agreed on \audit{agreed} of
\audit{judgments} judgments.  They found two systematic errors, which
two exclusion rules remove, and both judged all
\audit{overrides_sampled} sampled label corrections correct.

\textbf{Task audit.}  They also reviewed \taudit{families} families (all
\taudit{nvr} NVR and five satisfiable per tier) for unambiguous
predicates, consistent variants, faithful wording, no giveaway of a hidden value,
and NVR plausibility, agreeing on all \taudit{judgments} applicable judgments and
finding no defect.

\textbf{Reproducibility.}  Track \texttt{\art{track_id}} fixes the catalog, tasks, policy, harness, and user profiles.  A saved snapshot of \art{source_responses} Wikidata query responses
rebuilds the catalog, tasks, and answers byte for byte; released checks replay
every archived tool call, score, and rule-based user reply, and scripts regenerate
every result from the traces and labels without an LLM.

\section{Results}

\begin{figure}[tbp]
\centering
\definecolor{agentdeepseek}{HTML}{2A78D6}
\definecolor{agentglm}{HTML}{EB6834}
\definecolor{agentqwen}{HTML}{1BAF7A}
\definecolor{agentgpt}{HTML}{EDA100}
\begin{tikzpicture}
\begin{groupplot}[
  group style={group size=3 by 1,horizontal sep=0.25cm,y descriptions at=edge left},
  scale only axis,height=2.2cm,ymin=0,ymax=1.04,legend image post style={xscale=0.6},
  ytick={0,0.2,0.4,0.6,0.8,1},
  yticklabel style={/pgf/number format/fixed,/pgf/number format/fixed zerofill,/pgf/number format/precision=1},
  grid=major,grid style={gray!18},
  tick label style={font=\footnotesize},label style={font=\footnotesize},title style={font=\footnotesize},
  every axis plot/.append style={line width=0.8pt,mark size=2pt},
]
\nextgroupplot[title={(a) By condition},ylabel={$\passhat{1}$},width=3.65cm,xmin=0.5,xmax=3.45,xtick={1,2,3},xticklabels={Volunteer,Mixed,Hidden},legend to name=headroom-legend,legend style={draw=none,legend columns=6,font=\footnotesize,/tikz/every even column/.append style={column sep=0.35em}}]
\addplot[agentdeepseek,mark=*,error bars/.cd,y dir=both,y explicit] table[x=x,y=y,y error minus=m,y error plus=p,row sep=\\]{
x y m p\\
0.88 0.9944 0.0111 0.0056\\
1.88 0.9194 0.0444 0.0389\\
2.88 0.3278 0.0931 0.0972\\
};
\addlegendentry{DeepSeek}
\addplot[agentglm,mark=square*,error bars/.cd,y dir=both,y explicit] table[x=x,y=y,y error minus=m,y error plus=p,row sep=\\]{
x y m p\\
0.96 0.9889 0.0222 0.0111\\
1.96 0.9083 0.0361 0.0333\\
2.96 0.3458 0.0792 0.0833\\
};
\addlegendentry{GLM}
\addplot[agentqwen,mark=triangle*,error bars/.cd,y dir=both,y explicit] table[x=x,y=y,y error minus=m,y error plus=p,row sep=\\]{
x y m p\\
1.04 0.9208 0.0500 0.0444\\
2.04 0.7556 0.0653 0.0625\\
3.04 0.3097 0.0847 0.0861\\
};
\addlegendentry{Qwen}
\addplot[agentgpt,mark=diamond*,error bars/.cd,y dir=both,y explicit] table[x=x,y=y,y error minus=m,y error plus=p,row sep=\\]{
x y m p\\
1.12 0.9972 0.0056 0.0028\\
2.12 0.4389 0.0889 0.0903\\
3.12 0.1472 0.0667 0.0750\\
};
\addlegendentry{GPT-6.1 Sol}
\addplot[black,mark=o] coordinates {(1,1.0000) (2,1.0000) (3,0.9556)};
\addlegendentry{Eliminating}
\addplot[black!60,dashed,mark=none] coordinates {(1,1.0000) (2,1.0000) (3,0.6222)};
\addlegendentry{Popularity}
\nextgroupplot[title={(b) Hidden, by rejections},width=2.5cm,xmin=0,xmax=17,xtick={0,3,6,9,12,15}]
\addplot[agentdeepseek,const plot,mark=*,mark repeat=3] coordinates {(0,0.0986) (1,0.1944) (2,0.2694) (3,0.3069) (4,0.3264) (5,0.3264) (6,0.3264) (7,0.3264) (8,0.3264) (9,0.3264) (10,0.3264) (11,0.3264) (12,0.3278) (13,0.3278) (14,0.3278) (15,0.3278) (16,0.3278) (17,0.3278)};
\addplot[agentglm,const plot,mark=square*,mark repeat=3] coordinates {(0,0.0833) (1,0.1292) (2,0.1639) (3,0.1903) (4,0.2181) (5,0.2458) (6,0.2611) (7,0.2778) (8,0.2944) (9,0.3208) (10,0.3236) (11,0.3236) (12,0.3278) (13,0.3403) (14,0.3444) (15,0.3444) (16,0.3458) (17,0.3458)};
\addplot[agentqwen,const plot,mark=triangle*,mark repeat=3] coordinates {(0,0.0764) (1,0.1542) (2,0.1764) (3,0.1903) (4,0.2042) (5,0.2264) (6,0.2458) (7,0.2639) (8,0.2681) (9,0.2792) (10,0.2903) (11,0.2944) (12,0.2986) (13,0.3028) (14,0.3083) (15,0.3083) (16,0.3083) (17,0.3097)};
\addplot[agentgpt,const plot,mark=diamond*,mark repeat=3] coordinates {(0,0.0931) (1,0.1417) (2,0.1444) (3,0.1444) (4,0.1444) (5,0.1472) (6,0.1472) (7,0.1472) (8,0.1472) (9,0.1472) (10,0.1472) (11,0.1472) (12,0.1472) (13,0.1472) (14,0.1472) (15,0.1472) (16,0.1472) (17,0.1472)};
\addplot[black,mark=o,const plot,mark repeat=3] coordinates {(0,0.2444) (1,0.4000) (2,0.5556) (3,0.6222) (4,0.6444) (5,0.7333) (6,0.8222) (7,0.8667) (8,0.8667) (9,0.8889) (10,0.9111) (11,0.9111) (12,0.9333) (13,0.9333) (14,0.9333) (15,0.9333) (16,0.9333) (17,0.9556)};
\addplot[black!60,dashed,mark=none,const plot,mark repeat=3] coordinates {(0,0.1111) (1,0.1556) (2,0.2444) (3,0.3111) (4,0.3333) (5,0.3556) (6,0.4000) (7,0.4222) (8,0.4667) (9,0.4667) (10,0.4667) (11,0.4667) (12,0.4889) (13,0.5333) (14,0.5556) (15,0.5778) (16,0.6000) (17,0.6222)};
\nextgroupplot[title={(c) Hidden, by prompt},width=4.00cm,xmin=0.65,xmax=4.35,xtick={1,2,3,4},xticklabels={Base,Ask,Verify,Protocol}]
\addplot[agentdeepseek,mark=*,only marks,error bars/.cd,y dir=both,y explicit] table[x=x,y=y,y error minus=m,y error plus=p,row sep=\\]{
x y m p\\
0.88 0.3417 0.1833 0.1917\\
1.88 0.4417 0.1917 0.2000\\
2.88 0.4833 0.1833 0.1833\\
3.88 0.8167 0.1583 0.1250\\
};
\addplot[agentglm,mark=square*,only marks,error bars/.cd,y dir=both,y explicit] table[x=x,y=y,y error minus=m,y error plus=p,row sep=\\]{
x y m p\\
0.96 0.3000 0.1417 0.1667\\
1.96 0.2833 0.1500 0.1667\\
2.96 0.5417 0.2167 0.2167\\
3.96 0.7083 0.2000 0.1750\\
};
\addplot[agentqwen,mark=triangle*,only marks,error bars/.cd,y dir=both,y explicit] table[x=x,y=y,y error minus=m,y error plus=p,row sep=\\]{
x y m p\\
1.04 0.2750 0.1500 0.1667\\
2.04 0.3333 0.1750 0.1917\\
3.04 0.5250 0.2250 0.2250\\
4.04 0.7167 0.2000 0.1750\\
};
\addplot[agentgpt,mark=diamond*,only marks,error bars/.cd,y dir=both,y explicit] table[x=x,y=y,y error minus=m,y error plus=p,row sep=\\]{
x y m p\\
1.12 0.0667 0.0667 0.1333\\
2.12 0.1417 0.1333 0.1833\\
3.12 0.6167 0.2250 0.2083\\
4.12 0.8250 0.1667 0.1333\\
};
\addplot[black,mark=o,mark=none] coordinates {(0.65,1.0000) (4.35,1.0000)};
\addplot[black!60,dashed,mark=none,mark=none] coordinates {(0.65,0.6000) (4.35,0.6000)};
\end{groupplot}
\path (group c1r1.south west)--(group c3r1.south east) node[midway,below=4.5mm]{\ref{headroom-legend}};
\end{tikzpicture}
\caption{(a, b) The panel's 45 satisfiable families, agents pooled over interfaces (95\% family-bootstrap intervals), and the scripted agents on structured search; (b) counts a hidden success only within $r$ rejections.  (c) Hidden satisfiable success on the reliability block by prompt (structured search; prompts in Table~\ref{tab:usercard}).}
\label{fig:headroom}
\end{figure}
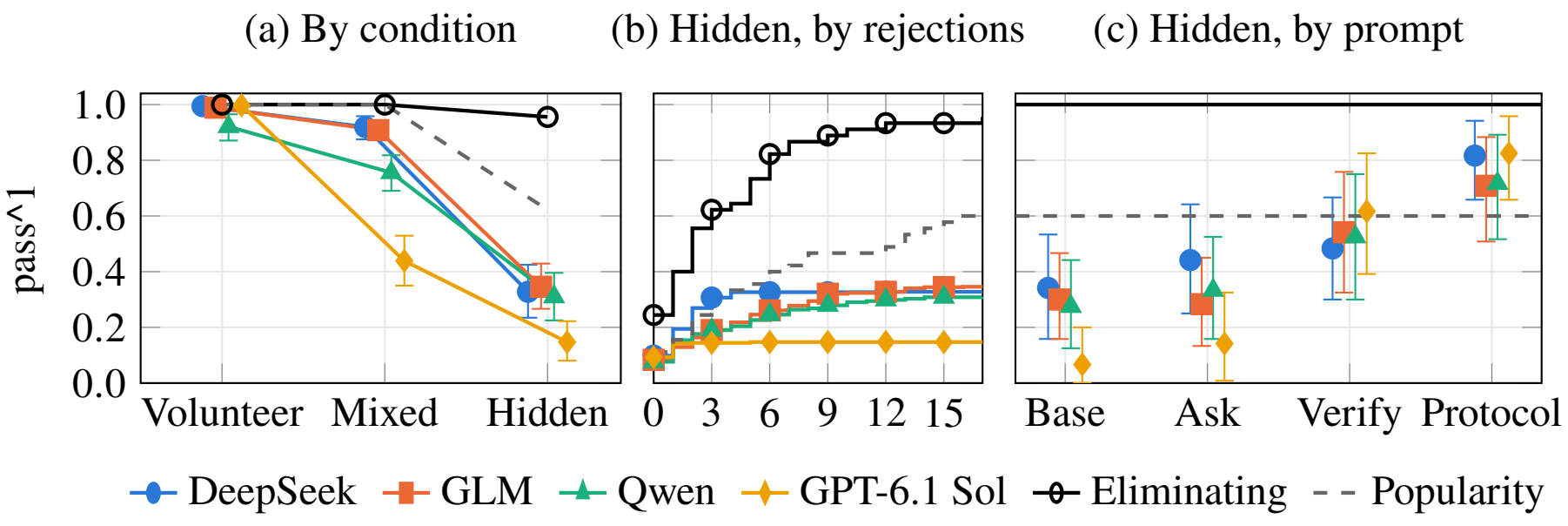

\subsection{RQ1: Constraint Disclosure}
\label{sec:rq1}

Success drops sharply when requirements are not stated up front, for every model
and interface (Table~\ref{tab:primary-interface}, Fig.~\ref{fig:headroom}a).
Averaged over models and interfaces, $\passhat{1}$ on
satisfiable families is \posthoc{volunteer_satisfiable} on volunteer,
\posthoc{mixed_satisfiable} on mixed, and \posthoc{hidden_satisfiable} on hidden
tasks.  Both gaps hold after Holm correction (Table~\ref{tab:contrasts}).  The mixed gap is
\pp{secondary/disclosure/mixed_minus_volunteer/pass^1}~points (all families) and the hidden gap, on satisfiable families,
\pp{secondary/disclosure/hidden_minus_volunteer/satisfiable/pass^1}~points.  Success also falls as tasks get
more complex, most on hidden tasks (\posthoc{hidden_simple} simple,
\posthoc{hidden_complex} complex).

\begin{table}[tbp]
\caption{$\passhat{k}$ on the panel at 20k.  \emph{Elic.}: share of mixed trials eliciting every askable fact.  LLM average: mean of the eight LLM rows.  Bold: best LLM agent per $\passhat{k}$ column.}
\label{tab:primary-interface}
\centering
\scriptsize
\setlength{\tabcolsep}{2.4pt}
\begin{tabular}{@{}ll ccc c cc c@{}}
\toprule
& & \multicolumn{3}{c}{Satisfiable (45 families)} & NVR (6) & \multicolumn{2}{c}{All (153 tasks)} & Mixed (51) \\
\cmidrule(lr){3-5}\cmidrule(lr){6-6}\cmidrule(lr){7-8}\cmidrule(lr){9-9}
Agent & Interface & Volunteer & Mixed & Hidden & Hidden & $\passhat{1}$ & $\passhat{4}$ & Elic. \\
\midrule
DeepSeek & keyword & 0.989 & 0.872 & 0.297 & 0.000 & 0.691 & 0.609 & 0.576 \\
 & structured & \textbf{1.000} & \textbf{0.967} & 0.358 & 0.000 & \textbf{0.760} & \textbf{0.666} & 0.859 \\
GLM & keyword & 0.978 & 0.869 & 0.325 & \textbf{0.167} & 0.714 & 0.574 & 0.569 \\
 & structured & \textbf{1.000} & 0.947 & \textbf{0.367} & 0.042 & 0.756 & 0.627 & 0.641 \\
Qwen & keyword & 0.842 & 0.569 & 0.294 & \textbf{0.167} & 0.550 & 0.381 & 0.377 \\
 & structured & \textbf{1.000} & 0.942 & 0.325 & 0.000 & 0.737 & 0.625 & 0.656 \\
GPT-6.1 Sol & keyword & 0.994 & 0.458 & 0.139 & 0.000 & 0.523 & 0.456 & 0.207 \\
 & structured & \textbf{1.000} & 0.419 & 0.156 & 0.000 & 0.511 & 0.424 & 0.159 \\
\addlinespace[2pt]
LLM average & both & 0.975 & 0.756 & 0.283 & 0.047 & -- & -- & -- \\
\midrule
Scripted, popularity & structured & 1.000 & 1.000 & 0.622 & 0.000 & 0.850 & 0.850 & -- \\
Scripted, eliminating & structured & 1.000 & 1.000 & 0.956 & 0.833 & 0.980 & 0.980 & -- \\
No-ask floor & -- & 1.000 & 0.067 & 0.067 & 0.000 & 0.373 & 0.373 & -- \\
\bottomrule
\end{tabular}
\end{table}

\begin{table}[tbp]
\caption{Paired $\passhat{1}$ contrasts in points, with 95\% family-bootstrap intervals.  Interface rows are structured minus keyword at 20k; the interaction is that difference at 5k minus at 20k.  $p$ values come from the family bootstrap and an exact sign-flip test over templates (9 for the LLM-user rows, 13 otherwise); $\le 0.0011$ is the smallest Holm-adjusted value 10{,}000 resamples allow.}
\label{tab:contrasts}
\centering
\scriptsize
\setlength{\tabcolsep}{3.3pt}
\begin{tabular}{@{}lrcrr@{}}
\toprule
& & \multicolumn{2}{c}{Families} & Templates \\
\cmidrule(lr){3-4}\cmidrule(l){5-5}
Contrast & \smash{$\Delta\passhat{1}$} & 95\% CI & $p$ & $p$ \\
\midrule
\multicolumn{5}{@{}l}{\emph{Tested alone}} \\
\quad Interface, model-averaged & $+7.1$ & $[+3.0,\,+11.4]$ & $0.0005$ & $0.15$ \\
\addlinespace[2pt]
\multicolumn{5}{@{}l}{\emph{Holm-corrected (11 tests)}} \\
\quad Interface, DeepSeek & $+6.9$ & $[+2.5,\,+11.6]$ & $0.0091$ & $0.38$ \\
\quad Interface, GLM & $+4.2$ & $[-0.7,\,+9.5]$ & $0.288$ & $0.69$ \\
\quad Interface, Qwen & $+18.7$ & $[+11.7,\,+25.9]$ & $\le 0.0011$ & $0.02$ \\
\quad Interface, GPT-6.1 Sol & $-1.2$ & $[-8.4,\,+6.0]$ & $0.753$ & $0.83$ \\
\quad Mixed $-$ volunteer (all families) & $-23.3$ & $[-27.3,\,-19.3]$ & $\le 0.0011$ & $0.003$ \\
\quad Hidden $-$ volunteer (satisfiable) & $-69.3$ & $[-76.3,\,-61.9]$ & $\le 0.0011$ & $0.005$ \\
\quad Keyword, 5k $-$ 20k & $+4.2$ & $[+1.3,\,+7.0]$ & $0.0255$ & $0.42$ \\
\quad Structured, 5k $-$ 20k & $+0.8$ & $[-0.2,\,+1.8]$ & $0.288$ & $0.62$ \\
\quad Interface $\times$ scale & $-3.3$ & $[-6.4,\,-0.3]$ & $0.140$ & $0.56$ \\
\quad GLM user $-$ rule-based & $-10.2$ & $[-15.7,\,-4.6]$ & $\le 0.0011$ & $0.38$ \\
\quad Qwen user $-$ rule-based & $-7.4$ & $[-13.2,\,-2.3]$ & $0.0138$ & $0.42$ \\
\bottomrule
\end{tabular}
\end{table}

\subsection{RQ2: Retrieval Interface at 20k}
\label{sec:rq2}

\textbf{Interface test.}  Structured search adds \pp{\primary}~points at 20k,
mostly through Qwen and on mixed tasks
(\pp{exploratory/interface/model_averaged/mixed/pass^1} points).  It holds across families (bootstrap $p\pvalr{\primary}$) but
not templates (exact sign-flip $p=\posthoc{signflip_template_primary}$), with or
without overflow and budget stops.  After Holm adjustment only Qwen's and
DeepSeek's gains hold;
without Qwen the average falls to
\pp{exploratory/interface/without_qwen38_27b_high/pass^1} points.  At 80\% power, the panel
detects effects of about \posthoc{mde_pp} points for this contrast and \posthoc{pairs_mde_pp}
between two agent--interface pairs (\posthoc{pairs_hidden_mde_pp} on hidden tasks).

\textbf{Tools or prompt.}  Adding the structured
tool's field list to the keyword prompt separates tool from prompt.  For DeepSeek the field list adds
\pp{\dsabl/field_list/all/pass^1} points and structured retrieval
\pp{\dsabl/retrieval_with_field_list/all/pass^1} more; for Qwen the field list
adds \pp{\qwabl/field_list/all/pass^1} and retrieval
\pp{\qwabl/retrieval_with_field_list/all/pass^1}.  Since every requirement is a
catalog field, this is a best case for filters.  Agents do not skip search by
recalling book IDs: only \summary{unseen_recommendations} of
\summary{final_recommendations} final recommendations name a book search never
returned, and all are invalid.

\subsection{RQ3: Catalog Scale (5k--20k)}

Catalog size matters far less than disclosure.  Answer sets grow with the catalog
(median \art{median_answers_5000} answers at 5k, \art{median_answers_20000} at
20k), and only the open-weight agents ran this study, on volunteer and mixed
tasks.  Fixed keyword queries find an answer for \retr{keyword_5000} of the 135
satisfiable families at 5k but only \retr{keyword_20000} at 20k.  Agents lose far
less but search more: keyword $\passhat{1}$ falls
\pp{secondary/scale/keyword/5k_minus_20k/pass^1} points from 5k to 20k, mostly
through Qwen, and structured $\passhat{1}$ by
\pp{secondary/scale/structured/5k_minus_20k/pass^1} (Table~\ref{tab:contrasts}).

\subsection{RQ4: Reliability across Models, Trials, and Users}
\label{sec:rq4}

The disclosure gradient holds for every model and interface, at $\passhat{4}$
(averaged, \lvl{disclosure/volunteer/pass^4} on volunteer,
\lvl{disclosure/mixed/pass^4} on mixed, and \lvl{disclosure/hidden/pass^4} on
hidden tasks), and under both LLM users.

\textbf{LLM users.}  With DeepSeek and structured search on the reliability
block, GLM and Qwen users lower $\passhat{1}$ by
\pp{secondary/simulator/sim_glm53_flash_minus_symbolic/pass^1} and
\pp{secondary/simulator/sim_qwen38_27b_minus_symbolic/pass^1} points, mostly on mixed tasks, stating
every askable value in about half the trials (rule-based user:
\userstat{symbolic}{mixed_all_askable_stated_tolerant}\%).  They rarely leak
hidden values ($\le$\userstat{sim_glm53_flash}{hidden_value_stated_tolerant}\%
of trials \cite{zhu2024reliable}); GPT-6.1 Sol, which asks less, changes at most
\pp{addendum/gpt_users/sim_glm53_flash_minus_symbolic/pass^1} points and trails
DeepSeek under every user.

\subsection{Agent Failure Modes}
\label{sec:diagnostics}

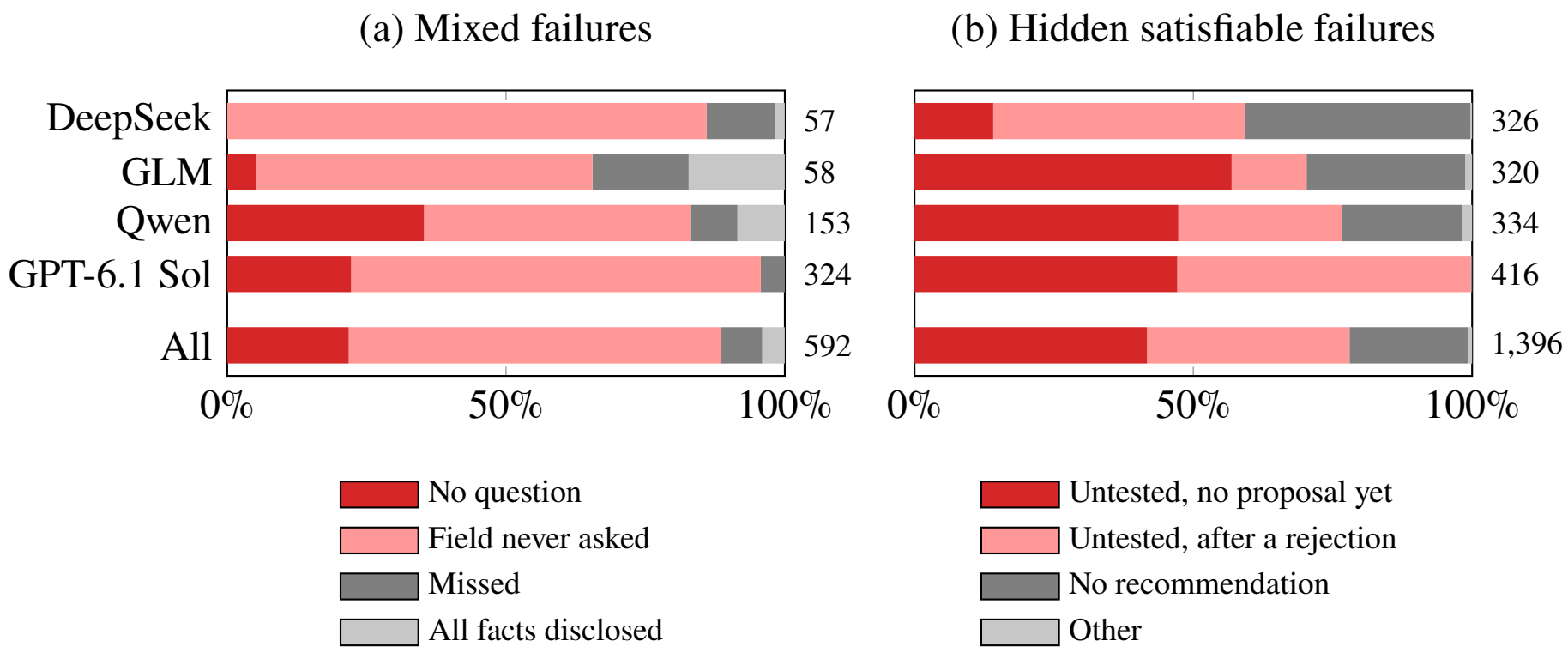
\begin{figure}[tbp]
\centering
\definecolor{fail0}{HTML}{D62728}
\definecolor{fail1}{HTML}{FF9896}
\definecolor{fail2}{HTML}{7F7F7F}
\definecolor{fail3}{HTML}{C7C7C7}
\begin{tikzpicture}
\begin{groupplot}[
  group style={group size=2 by 1,horizontal sep=1.0cm,y descriptions at=edge left},
  xbar stacked,scale only axis,width=4.3cm,height=2.2cm,/pgf/bar width=0.28cm,clip=false,
  xmin=0,xmax=1,xtick={0,0.5,1},xticklabels={0\%,50\%,100\%},
  ytick={1,2,3,4,5.4},yticklabels={DeepSeek,GLM,Qwen,GPT-6.1 Sol,All},y dir=reverse,ymin=0.4,ymax=6,
  tick label style={font=\footnotesize},title style={font=\footnotesize},major tick length=2pt,
  legend style={draw=none,font=\scriptsize,legend columns=1,at={(0.5,-0.3)},anchor=north},
  legend cell align=left,
]
\nextgroupplot[title={(a) Mixed failures}]
\addplot[fill=fail0,draw=none,area legend] coordinates {(0.0000,1) (0.0517,2) (0.3529,3) (0.2222,4) (0.2179,5.4)};
\addlegendentry{No question}
\addplot[fill=fail1,draw=none,area legend] coordinates {(0.8596,1) (0.6034,2) (0.4771,3) (0.7346,4) (0.6672,5.4)};
\addlegendentry{Field never asked}
\addplot[fill=fail2,draw=none,area legend] coordinates {(0.1228,1) (0.1724,2) (0.0850,3) (0.0432,4) (0.0743,5.4)};
\addlegendentry{Missed}
\addplot[fill=fail3,draw=none,area legend] coordinates {(0.0175,1) (0.1724,2) (0.0850,3) (0.0000,4) (0.0405,5.4)};
\addlegendentry{All facts disclosed}
\node[anchor=west,font=\scriptsize,xshift=1pt] at (axis cs:1,1) {57};
\node[anchor=west,font=\scriptsize,xshift=1pt] at (axis cs:1,2) {58};
\node[anchor=west,font=\scriptsize,xshift=1pt] at (axis cs:1,3) {153};
\node[anchor=west,font=\scriptsize,xshift=1pt] at (axis cs:1,4) {324};
\node[anchor=west,font=\scriptsize,xshift=1pt] at (axis cs:1,5.4) {592};
\nextgroupplot[title={(b) Hidden satisfiable failures}]
\addplot[fill=fail0,draw=none,area legend] coordinates {(0.1411,1) (0.5687,2) (0.4731,3) (0.4712,4) (0.4169,5.4)};
\addlegendentry{Untested, no proposal yet}
\addplot[fill=fail1,draw=none,area legend] coordinates {(0.4509,1) (0.1344,2) (0.2934,3) (0.5264,4) (0.3632,5.4)};
\addlegendentry{Untested, after a rejection}
\addplot[fill=fail2,draw=none,area legend] coordinates {(0.4049,1) (0.2844,2) (0.2156,3) (0.0024,4) (0.2120,5.4)};
\addlegendentry{No recommendation}
\addplot[fill=fail3,draw=none,area legend] coordinates {(0.0031,1) (0.0125,2) (0.0180,3) (0.0000,4) (0.0079,5.4)};
\addlegendentry{Other}
\node[anchor=west,font=\scriptsize,xshift=1pt] at (axis cs:1,1) {326};
\node[anchor=west,font=\scriptsize,xshift=1pt] at (axis cs:1,2) {320};
\node[anchor=west,font=\scriptsize,xshift=1pt] at (axis cs:1,3) {334};
\node[anchor=west,font=\scriptsize,xshift=1pt] at (axis cs:1,4) {416};
\node[anchor=west,font=\scriptsize,xshift=1pt] at (axis cs:1,5.4) {1,396};
\end{groupplot}
\end{tikzpicture}
\caption{How failed trials end on the panel at 20k, as shares of each agent's failures over both interfaces ($n$ at the bar end; \emph{All} pools the four agents).  Missed: keyword matching missed a field question that LLM labels found; untested: never proposed for a verdict.}
\label{fig:failures}
\end{figure}

\textbf{Mixed tasks fail when agents do not ask.}  GPT-6.1 Sol nearly solves
volunteer tasks but has the lowest mixed $\passhat{1}$; in
\summary{gpt_no_elicitation_lo}--\summary{gpt_no_elicitation_hi}\% of its mixed
trials it elicits no askable fact (open-weight: at most
\summary{open_no_elicitation_hi}\%), and its \summary{gpt_complete_n} mixed trials
that elicit every askable fact all succeed.  Most failed mixed trials ask no question or never ask about some field
(Fig.~\ref{fig:failures}); trials eliciting every askable fact succeed
\posthoc{mixed_complete_success_pct}\% of the time, others
\posthoc{mixed_incomplete_success_pct}\%, since popular books often meet unasked
requirements.
On the reliability block, the ask prompt (Table~\ref{tab:usercard})
raises GPT-6.1 Sol's mixed $\passhat{1}$ by \pp{\gptvariant/mixed/pass^1} points
and lifts every agent to at least \headroom{ask_mixed_lo}.  Keyword-matching errors, measured with LLM labels,
move mixed success by at most \misread{drop_all_pp} points down or \missq{labelled_pp} up.

\textbf{Hidden tasks fail on untested commitments.}  In \hf{untested_pct}\% of
failed hidden satisfiable trials, the agent recommends a book it never proposed
for a verdict (GPT-6.1 Sol: \hf{gpt_untested} of \hf{gpt_failures}), half of
them before any proposal; most (\hf{refused_untested_pct}\%) follow a
declined question, which signals a hidden requirement.  Few come from the tag
format: in \hf{untagged_acceptable_pct}\% of failures an earlier turn had named an
acceptable book without a tag.  Most other failures end at
the time or turn cap.

\textbf{How far proposing and elimination go.}  The popularity agent solves
\scripted{hidden_sat_panel_solved} of the \scripted{hidden_sat_panel} satisfiable
hidden tasks, the eliminating agent \elim{hidden_sat_panel_solved}; within three
rejections they reach \posthoc{capped3_scripted} and
\posthoc{capped3_eliminating}, agents \posthoc{capped3_all}
(Fig.~\ref{fig:headroom}b).  Told to propose first (ask-and-verify,
Fig.~\ref{fig:headroom}c), agents cut untested commitments to at most
\headroom{verify_untested_hi_pct}\% of hidden failures and reach
\headroom{verify_sat_lo}--\headroom{verify_sat_hi} on the reliability block, not
detectably different from the popularity agent's \scripted{hidden_satisfiable_rel}
there.

\textbf{What the protocol adds.}  The eliminating
agent (\elim{hidden_satisfiable_rel}) holds three things the agents lack: the
meaning of a decline and of a rejection, a list of possible hidden requirements,
and systematic elimination.  The protocol prompt
(Table~\ref{tab:usercard}) supplies the first.  It lifts every agent, by
\headroom{protocol_gain_pp} points on average, to \headroom{protocol_sat_lo}--\headroom{protocol_sat_hi}, detectably above the
popularity agent for \headroom{protocol_vs_popularity_above} of the 4, and agents now abstain on
\headroom{protocol_nvr_lo_pct}--\headroom{protocol_nvr_hi_pct}\% of hidden NVR tasks
(ask-and-verify: at most \headroom{verify_nvr_hi_pct}\%).  All four stay below the
eliminating agent, also within three rejections
(\headroom{protocol_capped_lo}--\headroom{protocol_capped_hi} against
\posthoc{capped3_eliminating_rel}).  Knowing the protocol thus closes part of the
hidden gap, and the rest lies in search.  Every structured run
abstains on volunteer NVR tasks, but under the base prompt agents reach only \summary{hidden_nvr} on
hidden ones, where verdicts cannot prove that no book fits; the eliminating agent
abstains on \elim{hidden_nvr_panel_solved} of \elim{hidden_nvr_panel} once that
list runs out.

\section{Availability, Intended Use, and Maintenance}
\label{sec:avail}

\textbf{Availability.}  The catalog and its source snapshot, the tasks and answers, the
environment, validators, reference agents, and every trace are released (code MIT; catalog and snapshot
CC0~1.0; tasks, answers, and policies CC~BY~4.0) at
{\def\UrlBreaks{\do\/}\url{https://github.com/nbharaths/tau-rec/tree/main/data/tracks/books-wikidata-v2}}.

\textbf{Using the benchmark.}  To report results, run the panel at 20k
with the rule-based user, the 20-turn cap, at least four trials per task, one or
both interfaces, and two prompts: the base prompt, which measures whether an agent
asks and verifies unprompted, and ask-and-verify, under which hidden tasks measure
what remains once agents propose before committing.  Both prompts matter: GPT-6.1
Sol trails the other agents under the base prompt and catches up under
ask-and-verify.  Report $\passhat{1}$ by condition (hidden NVR tasks separately),
$\passhat{4}$, hidden success by rejections allowed, elicitation, how trials end,
and the provider.  One agent's structured run costs about
US\$\posthoc{run_usd_lo}--\posthoc{run_usd_hi} per prompt.  The panel separates
\posthoc{pairs_separated} of the \posthoc{pairs_total} agent--interface pairs in
Table~\ref{tab:primary-interface} at the 95\% level, and if it leaks into
training data, the generator can draw fresh families.

\textbf{Maintenance.}  Track IDs never change; fixes ship under a new ID, and the first author maintains
the resource through the issue tracker.

\section{Bias, Ethics, and Limitations}
\label{sec:limits}

\textbf{Catalog coverage.}  Wikidata and the English-title requirement favor
well-documented, Anglo-American works (English on \art{english_pct}\% of records,
the US on \art{us_pct}\%, the UK on \art{uk_pct}\%), and
\art{country_predicates_us_uk} of \art{country_predicates} task-country predicates
target the US or UK.  Structured search concentrates exposure on the most
popular books in each answer set, far more than keyword search.  Genres are
skewed toward science fiction, and literary fiction is absent. Future work can extend the catalog beyond English-titled works.

\textbf{Scope.}  The tasks are hard-filter elicitation over four metadata
fields; soft preferences, read-alikes, and taste, which real book requests often
contain \cite{koolen2015looking}, are out of scope.  The rule-based user speaks in
catalog labels and never explains a rejection, so the benchmark measures how
reliably agents elicit and verify requirements rather than conversational skill.
Mixed tasks test whether an agent asks, not how well, since one broad question
discloses every askable fact.  The user model has not been validated against
people, and we test one frontier model.

\section{Conclusion}
\label{sec:conclusion}

Holding each request fixed, \taurec{}-Books shows that when requirements are
revealed matters far more than how agents search or how large the catalog is.
Success on satisfiable tasks falls from \posthoc{volunteer_satisfiable} to
\posthoc{mixed_satisfiable} when requirements must be asked for and to
\posthoc{hidden_satisfiable} when some are never stated (RQ1).  Told to ask,
every agent reaches at least \headroom{ask_mixed_lo} on mixed tasks, but on hidden
tasks agents told to propose first stay well below an agent that rules out
requirements from rejections; explaining declines and rejections closes part of that gap.  Structured search adds \pp{\primary} points, mostly through one agent (RQ2);
a four times larger catalog costs keyword search
\pp{secondary/scale/keyword/5k_minus_20k/pass^1} points and structured search
almost nothing (RQ3); and the gradient holds for every model and interface, at
$\passhat{4}$, and with LLM users (RQ4).  Critiquing users, costlier
elicitation, soft preferences, and human-validated user models come next.

\begin{credits}
\subsubsection{\discintname}
\label{sec:disc}
The authors have no competing interests to declare that are relevant to the
content of this article.
\end{credits}

\bibliographystyle{splncs04}
\bibliography{references}

\end{document}